\documentclass{aastex63}
\usepackage{natbib}
\usepackage{graphicx}
\usepackage{graphics,epsfig}
\let\splitbox\relax
\usepackage{tabularx}
\usepackage{adjustbox}
\usepackage{lipsum}
\usepackage{amsmath}
\usepackage{float}
\usepackage{courier}
\usepackage{multirow}
\usepackage{xcolor}
\usepackage{url}
\usepackage{booktabs}
\usepackage{gensymb}
\usepackage[normalem]{ulem}
\usepackage{colortbl}
\usepackage{hyperref}
\hypersetup{colorlinks=true,
linkcolor=red,
citecolor=blue}
\usepackage{rotating}

\begin{document}
\let\linenumbers\relax
\title{The Atacama Cosmology Telescope: Summary of DR4 and DR5 Data Products and Data Access}

\author[0000-0002-2018-3807]{Maya~Mallaby-Kay}
\affiliation{Department of Astronomy, University of Chicago, Chicago, IL USA}

\author[0000-0002-2287-1603]{Zachary~Atkins}
\affiliation{Joseph Henry Laboratories of Physics, Jadwin Hall, Princeton University, Princeton, NJ, USA 08544}

\author[0000-0002-1035-1854]{Simone~Aiola}
\affiliation{Center for Computational Astrophysics, Flatiron Institute, 162 5th Avenue, New York, NY 10010 USA}

\author[0000-0002-4200-9965]{Stefania~Amodeo}
\affiliation{Department of Astronomy, Cornell University, Ithaca, NY 14853, USA}

\author{Jason~E.~Austermann}
\affiliation{NIST Quantum Devices Group, 325 Broadway Mailcode 817.03, Boulder, CO, USA 80305}

\author{James~A.~Beall}
\affiliation{NIST Quantum Devices Group, 325 Broadway Mailcode 817.03, Boulder, CO, USA 80305}

\author{Daniel~T.~Becker}
\affiliation{NIST Quantum Devices Group, 325 Broadway Mailcode 817.03, Boulder, CO, USA 80305}

\author[0000-0003-2358-9949]{J.~Richard~Bond}
\affiliation{Canadian Institute for Theoretical Astrophysics, 60 St. George Street,
University of Toronto, Toronto, ON, M5S 3H8, Canada}

\author[0000-0003-0837-0068]{Erminia~Calabrese}
\affiliation{School of Physics and Astronomy, Cardiff University, The Parade, Cardiff, CF24 3AA, UK}

\author{Grace~E.~Chesmore}
\affiliation{Department of Physics, University of Chicago, Chicago, IL 60637, USA} 
\author{Steve~K.~Choi}
\affiliation{Department of Physics, Cornell University, Ithaca, NY, USA 14853}
\affiliation{Department of Astronomy, Cornell University, Ithaca, NY 14853, USA}

\author{Kevin~T.~Crowley}
\affiliation{Department of Physics, University of California, Berkeley, LeConte Hall, Berkeley, CA 94720}

\author{Omar~Darwish}
\affiliation{Department of Applied Mathematics and Theoretical Physics, University of Cambridge, Wilberforce Road, Cambridge CB3 0WA, UK}

\author{Edward~V.~Denison}
\affiliation{NIST Quantum Devices Group, 325 Broadway Mailcode 817.03, Boulder, CO, USA 80305}

\author{Mark~J.~Devlin}
\affiliation{Department of Physics and Astronomy, University of Pennsylvania, 209 South 33rd Street, Philadelphia, PA, USA 19104}

\author{Shannon~M.~Duff}
\affiliation{NIST Quantum Devices Group, 325 Broadway Mailcode 817.03, Boulder, CO, USA 80305}

\author[0000-0003-2856-2382]{Adriaan~J.~Duivenvoorden}
\affiliation{Joseph Henry Laboratories of Physics, Jadwin Hall, Princeton University, Princeton, NJ, USA 08544}

\author{Jo~Dunkley}
\affiliation{Joseph Henry Laboratories of Physics, Jadwin Hall, Princeton University, Princeton, NJ, USA 08544}
\affiliation{Department of Astrophysical Sciences, Peyton Hall, Princeton University, Princeton, NJ USA 08544}

\author{Simone~Ferraro}
\affiliation{Lawrence Berkeley National Laboratory, One Cyclotron Road, Berkeley, CA 94720, USA}

\author{Kyra~Fichman}
\affiliation{Department of Physics, University of Michigan, Ann Arbor, USA 48103}

\author[0000-0001-9731-3617]{Patricio~A.~Gallardo}
\affiliation{Department of Physics, Cornell University, Ithaca, NY, USA 14853}

\author{Joseph~E.~Golec} 
\affiliation{Department of Physics, University of Chicago, Chicago, IL 60637, USA}

\author[0000-0002-1697-3080]{Yilun~Guan}
\affiliation{Department of Physics and Astronomy, University of Pittsburgh, Pittsburgh, PA, USA 15260}

\author{Dongwon~Han}
\affiliation{Physics and Astronomy Department, Stony Brook University, Stony Brook, NY 11794, USA}

\author{Matthew~Hasselfield}
\affiliation{Center for Computational Astrophysics, Flatiron Institute, 162 5th Avenue, New York, NY 10010 USA}

\author[0000-0002-9539-0835]{J.~Colin~Hill}
\affiliation{Department of Physics, Columbia University, 538 West 120th Street, New York, NY, USA 10027}
\affiliation{Center for Computational Astrophysics, Flatiron Institute, New York, NY, USA 10003}

\author{Gene~C.~Hilton}
\affiliation{NIST Quantum Devices Group, 325 Broadway Mailcode 817.03, Boulder, CO, USA 80305}

\author[0000-0002-8490-8117]{Matt~Hilton}
\affiliation{Astrophysics Research Centre, University of KwaZulu-Natal, Westville Campus, Durban 4041, South Africa}
\affiliation{School of Mathematics, Statistics \& Computer Science, University of KwaZulu-Natal, Westville Campus, Durban 4041, South Africa}

\author{Ren\'ee~Hlo\v{z}ek}
\affiliation{David A. Dunlap Department of Astronomy and Astrophysics, 50 St. George Street, Toronto ON M5S3H4}
\affiliation{Dunlap Institute for Astronomy and Astrophysics, 50 St. George Street, Toronto ON M5S3H4}

\author{Johannes~Hubmayr}
\affiliation{NIST Quantum Devices Group, 325 Broadway Mailcode 817.03, Boulder, CO, USA 80305}

\author[0000-0001-7109-0099]{Kevin~M.~Huffenberger}
\affiliation{Department of Physics, Florida State University, Tallahassee, FL 32306, USA}

\author[0000-0002-8816-6800]{John~P.~Hughes}
\affiliation{Department of Physics and Astronomy, Rutgers, the State University of New Jersey, 136 Frelinghuysen Road, Piscataway, NJ 08854-8019, USA}

\author{Brian~J.~Koopman}
\affiliation{Department of Physics, Yale University, New Haven, CT 06520, USA}

\author{Thibaut~Louis}
\affiliation{Universit\'e Paris-Saclay, CNRS/IN2P3, IJCLab, 91405 Orsay, France}

\author{Amanda~MacInnis}
\affiliation{Physics and Astronomy Department, Stony Brook University, Stony Brook, NY 11794, USA}

\author{Mathew~S.~Madhavacheril}
\affiliation{Centre for the Universe, Perimeter Institute, Waterloo, ON N2L 2Y5, Canada}

\author{Jeff~McMahon}
\affiliation{Department of Astronomy, University of Chicago, Chicago, IL USA}
\affiliation{Department of Physics, University of Chicago, Chicago, IL 60637, USA}
\affiliation{Kavli Institute for Cosmological Physics, University of Chicago, 5640 S. Ellis Ave., Chicago, IL 60637, USA19}
\affiliation{Enrico Fermi Institute, University of Chicago, Chicago, IL 60637, USA}

\author{Kavilan~Moodley}
\affiliation{Astrophysics Research Centre, University of KwaZulu-Natal, Westville Campus, Durban 4041, South Africa}
\affiliation{School of Mathematics, Statistics \& Computer Science, University of KwaZulu-Natal, Westville Campus, Durban 4041, South Africa}

\author[0000-0002-4478-7111]{Sigurd~Naess}
\affiliation{Center for Computational Astrophysics, Flatiron Institute, 162 5th Avenue, New York, NY 10010 USA}

\author{Toshiya~Namikawa}
\affiliation{Department of Applied Mathematics and Theoretical Physics, University of Cambridge, Wilberforce Road, Cambridge CB3 0WA, United Kingdom}

\author[0000-0002-8307-5088]{Federico~Nati}
\affiliation{Department of Physics, University of Milano-Bicocca, Piazza della Scienza 3, 20126 Milano (MI), Italy}

\author{Laura~B.~Newburgh}
\affiliation{Department of Physics, Yale University, New Haven, CT 06520, USA}

\author{John~P.~Nibarger}
\affiliation{NIST Quantum Devices Group, 325 Broadway Mailcode 817.03, Boulder, CO, USA 80305}

\author[0000-0001-7125-3580]{Michael~D.~Niemack}
\affiliation{Department of Astronomy, Cornell University, Ithaca, NY 14853, USA}
\affiliation{Department of Physics, Cornell University, Ithaca, NY, USA 14853}

\author[0000-0002-9828-3525]{Lyman~A.~Page}
\affiliation{Joseph Henry Laboratories of Physics, Jadwin Hall, Princeton University, Princeton, NJ, USA 08544}

\author[0000-0003-4006-1134]{Maria~Salatino} 
\affiliation{Kavli Institute for Particle Astrophysics and Cosmology, Stanford, CA 94305 USA}
\affiliation{Department of Physics, Stanford University, Stanford, CA 94305, USA, Stanford, CA 94305 USA} 

\author{Emmanuel~Schaan}
\affiliation{Lawrence Berkeley National Laboratory, One Cyclotron Road, Berkeley, CA 94720, USA}
\affiliation{Berkeley Center for Cosmological Physics, UC Berkeley, CA 94720, USA}

\author{Alessandro~Schillaci}
\affiliation{Department of Physics, California Institute of Technology, Pasadena, CA91125, USA.}

\author{Neelima~Sehgal}
\affiliation{Physics and Astronomy Department, Stony Brook University, Stony Brook, NY 11794, USA}

\author{Blake~D.~Sherwin}
\affiliation{Department of Applied Mathematics and Theoretical Physics, University of Cambridge, Wilberforce Road, Cambridge CB3 0WA, UK}

\author[0000-0002-8149-1352]{Crist\'obal~Sif\'on}
\affiliation{Instituto de F\'isica, Pontificia Universidad Cat\'olica de Valpara\'iso, Casilla 4059, Valpara\'iso, Chile}

\author{Sara~Simon}
\affiliation{Department of Physics, University of Michigan, Ann Arbor,
USA 48103}

\author[0000-0002-7020-7301]{Suzanne~T.~Staggs}
\affiliation{Joseph Henry Laboratories of Physics, Jadwin Hall, Princeton University, Princeton, NJ, USA 08544}

\author{Emilie~R.~Storer}
\affiliation{Joseph Henry Laboratories of Physics, Jadwin Hall, Princeton University, Princeton, NJ, USA 08544}

\author{Joel~N.~Ullom}
\affiliation{NIST Quantum Devices Group, 325 Broadway Mailcode 817.03, Boulder, CO, USA 80305}

\author{Alexander~Van~Engelen}
\affiliation{School of Earth and Space Exploration, Arizona State University, Tempe, AZ 85287, USA}

\author{Jeff~Van~Lanen}
\affiliation{NIST Quantum Devices Group, 325 Broadway Mailcode 817.03, Boulder, CO, USA 80305}

\author{Leila~R.~Vale}
\affiliation{NIST Quantum Devices Group, 325 Broadway Mailcode 817.03, Boulder, CO, USA 80305}

\author[0000-0002-7567-4451]{Edward~J.~Wollack}
\affiliation{NASA/Goddard Space Flight Center, Greenbelt, MD 20771, USA}

\author[0000-0001-5112-2567]{Zhilei~Xu}
\affiliation{Department of Physics and Astronomy, University of Pennsylvania, 209 South 33rd Street, Philadelphia, PA, USA 19104}
\affiliation{MIT Kavli Institute, Massachusetts Institute of Technology, Cambridge, MA, USA 02139}

\begin{abstract}
Two recent large data releases for the Atacama Cosmology Telescope (ACT), called DR4 and DR5, are available for public access. These data include temperature and polarization maps that cover nearly half the sky at arcminute resolution in three frequency bands; lensing maps and component-separated maps covering $\sim$ 2,100 deg$^2$ of sky; derived power spectra and cosmological likelihoods; a catalog of over 4,000 galaxy clusters; and supporting ancillary products including beam functions and masks. The data and products are described in a suite of ACT papers; here we provide a summary. In order to facilitate ease of access to these data we present a set of Jupyter IPython notebooks developed to introduce users to DR4, DR5, and the tools needed to analyze these data. The data products (excluding simulations) and the set of notebooks are publicly available on the NASA Legacy Archive for Microwave Background Data Analysis (LAMBDA); simulation products are available on the National Energy Research Scientific Computing Center (NERSC).

\end{abstract}

\section{Introduction\label{sec:intro}}
The Atacama Cosmology Telescope (ACT) is a 6\,m telescope in the Atacama Desert in northern Chile with arcminute resolution \citep{Fowler:07,Thornton2016}.  It has been mapping the millimeter sky since 2007, with three successive cameras. This paper summarizes the fourth and fifth data releases from ACT, referred to as DR4 and DR5. All the elements of the data releases are described in individual papers as outlined below; this paper gives a summary of the contents of the two data releases and describes a suite of Jupyter IPython notebooks that accompany the data products to facilitate their use.

The majority of the DR4 products use data collected from 2013--16 (hereafter s13--s16, likewise for other data seasons). These data cover $\sim$ 18,000\,deg$^2$, resulting in maps made in seven regions of the sky, in frequency bands centered at 98 and 150\,GHz in both temperature and polarization. DR4 also includes data collected from 2008--10 (s08--s10). The maps from s08--s10 and s13--s16 are combined with those from the {\sl Planck} satellite to create high-resolution ``coadded" maps over the full footprint.
The DR4 data products and pipelines are described in more detail in ACT collaboration publications as follows. The individual s13--s16 maps are described in \cite{aiola/etal:2020} (A20, hereafter), while \cite{choi/etal:2020} (C20, hereafter) provides additional details on their validation. The coadded s08--16 maps with {\sl Planck} are in \cite{naess2020} (N20, hereafter). Derived CMB lensing maps are presented in \cite{darwish2020atacama}, and component separated CMB and Compton $y$-parameter maps in \cite{madhavacheril2019atacama}. The angular power spectra are described in C20, and cosmological likelihoods and parameters in A20. Delensed CMB power spectra are reported in \cite{hanetal2020}, and the birefringence spectra in \cite{Namikawa2020}.

DR5 adds ACT data acquired in 2017--18 (s17--s18) in frequency bands centered at 98, 150 and 220\,GHz. The DR5 data products  released so far consist of the coadded s08--s18 maps described in N20, and the derived galaxy cluster catalog presented in \cite{hilton_atacama_2020}. 

The outline of this paper is as follows. In \S \ref{sec: ACT} we describe the ACT instrument and the scope of DR4 and DR5. In \S \ref{sec: data} we describe the individual data products that are publicly available on the NASA Legacy Archive for Microwave Background Data Analysis\footnote{\url{https://lambda.gsfc.nasa.gov/product/act/actpol_prod_table.cfm}} (LAMBDA) and at the National Energy Research Scientific Computing Center (NERSC). In \S \ref{sec:notebooks} we give an introduction to the Jupyter IPython notebooks, which contain examples of working with DR4 and DR5 data as well as explanations of how to perform certain analyses. 
We summarize in \S \ref{sec: conc}. While this paper provides a summary,
any work using these data products or notebooks should reference the appropriate science publication(s) \citep{aiola/etal:2020, choi/etal:2020, naess2020, darwish2020atacama, madhavacheril2019atacama, hanetal2020, Namikawa2020, hilton_atacama_2020}.

\newpage

\begin{table*}
\centering
\setlength{\tabcolsep}{8pt} 
\renewcommand{\arraystretch}{1.00}
\begin{adjustbox}{width=\textwidth}
\movetableright=-0.6in
\begin{tabular}{l|l|l|l}
\multicolumn{4}{c}{\normalsize{Summary of DR4 and DR5 Data Products}}\\\addlinespace[7pt]
\toprule
    \textbf{Category} & \textbf{Dataset} & \textbf{Products} &\textbf{Suggested Uses}\\
\midrule
    \multicolumn{4}{l}{\textbf{DR4 Data Products}}\\
\midrule 
    \multirow{4}{*}{Frequency Maps} & \multirow{4}{*}{Per season/patch/array/frequency} & Source-free maps and splits &  Precision cosmology (e.g. see\\
    & & Source maps for above & notebooks 8 \S \ref{sec: nb8} and 9 \\
    & & Inverse-variance maps for above & \S \ref{sec: nb_9})\\
    & & Cross-linking maps for above &\\
\midrule
    \multirow{8}{*}{Derived Maps}
    & \multirow{3}{*}{Coadd maps of s08--s16 data} & Source-free maps & Studying arcminute-scales\\
    & & Source maps for above & Studying galaxy clusters, point\\
    & & Inverse-variance maps for above & sources, and Galactic emission\\
    \cmidrule{2-4}
    & \multirow{2}{*}{Component separated maps} & CMB+kSZ maps &\\
    & & tSZ/Compton-$y$ maps & Cross correlations with external \\
    \cmidrule{2-3}
    & \multirow{3}{*}{Lensing maps} & \multirow{3}{*}{Lensing-$\kappa$ maps} & data (e.g. see notebook 5\\
    & & & \S \ref{sec: nb5})\\
    & & &\\
\midrule
    \multirow{8}{*}{Ancillary Products} & \multirow{4}{*}{Masks} & Footprint masks\\
    & & Point source masks &\\
    & & Lensing masks &\\
    & & Masks for component separated maps & Needed for a range of \\ \cmidrule{2-3}
    & \multirow{4}{*}{Window functions} &  Beams & cosmological analyses\\
    & & Leakage beams\\
    & & Beams for component separated maps\\
    & & Map-making transfer functions\\ 
\midrule
    \multirow{4}{*}{Derived Spectra} & & Multifrequency spectra &\\
    & & CMB-only spectra & Plotting binned spectra\\
    & & Birefringence spectra & Comparisons to other data sets\\
    & & Delensed spectra &\\
    \midrule
    \multirow{3}{*}{Likelihoods} & Multifrequency likelihood & Fortran implementation & Testing theoretical models\\
    \cmidrule{2-3}
    & Delensed likelihood & CosmoMC compatible code & Combining with other data sets\\ 
    \cmidrule{2-3}
    & CMB-only likelihood & Fortran and Python implementations & Examining covariance matrices\\
\midrule
    \multirow{7}{*}{Cosmological Results} & \multirow{2}{*}{Likelihood products} & Best fit parameters &\\
    & & Chains of cosmological parameters & Comparing with other data sets\\
    \cmidrule{2-3}
    & \multirow{4}{*}{Delensing likelihood products} & Best fit parameters & Plotting parameter distributions\\
    & & Delensed theory & (e.g. see notebook 11 \S \ref{sec: nb11})\\
    & & Chains of cosmological parameters\\
    & & Parameter-shift covariance matrix
\\
\midrule
    \multirow{5}{*}{Simulations} & \multirow{2}{*}{Maps} & Signal simulations & These can be used to\\
    & & Lensing kappa simulations  &  validate pipelines and\\
    \cmidrule{2-3}
    & \multirow{3}{*}{2D Power spectra} & Square-root covariance matrices & for error analysis\\
    & & Lensed CMB $a_{lm}$ & (e.g. see notebook 9 \S \ref{sec: nb_9})\\
    & & Lensing phi $a_{lm}$ &\\
\midrule
    \multicolumn{4}{l}{\textbf{DR5 Data Products (Available to Date)}}\\ 
\midrule
    \multirow{3}{*}{Cluster Catalog} & & & Cluster cosmology, stacking\\
    & & 4,195 SZ clusters & analyses (e.g. see notebook 4\\ 
    & & &\S \ref{sec: nb4})\\
\midrule
    \multirow{4}{*}{Derived Maps}
    & \multirow{4}{*}{Coadd maps of s08--s18 data} & Source-free maps & Studying arcminute-scales\\
    & & Source maps for above & Studying galaxy clusters, point\\
    & & Inverse-variance maps for above & sources, and Galactic emission\\
    & & &  (e.g. see notebook 2 \S \ref{sec: nb2})\\
\bottomrule
\end{tabular}
\end{adjustbox}
\caption{\label{table:data_products} {\small An outline of the data products included in the two data releases, DR4 and DR5. Information relating to the particular products is discussed in \S \ref{sec: data}. The last column gives some suggested applications, with examples discussed in \S \ref{sec: notebooks}. The majority of data products are DR4 (using s08--s16 data); the s08--s18 coadd maps and derived cluster catalog belong to DR5.} }
\end{table*}

\section{The Atacama Cosmology Telescope}\label{sec: ACT}
ACT comprises a 6\,m primary and a 2\,m secondary in an off-axis, Gregorian telescope design \citep{Fowler:07}. ACT's elevation of $\sim$ 5,200\,m and the climate at its site on Cerro Toco in the Atacama Desert in Chile provide for excellent transparency at millimeter wavelengths.

\begin{figure}
    \centering
    \includegraphics[width=0.875\textwidth]{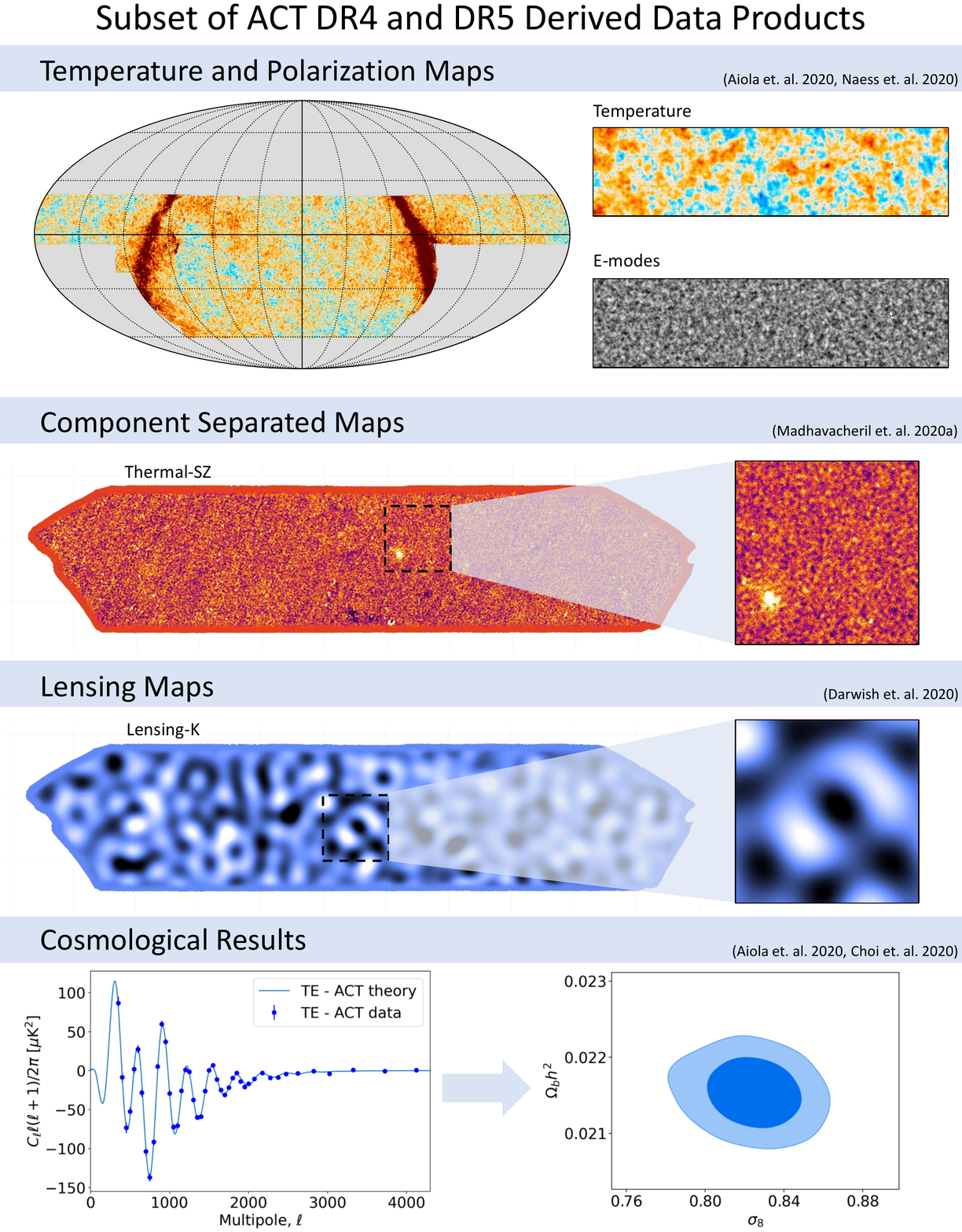}
    \caption{ACT DR4 and DR5 products available to date consist of more than 5\,TB of data including temperature and polarization maps, likelihoods, delensed spectra, and simulations.  Here we show a sample of the  products.  Top row:  coadded temperature and polarization maps of ACT with {\it Planck} (N20). The cutouts are centered at -162\degree\ (7\degree) with a dimension of 24\degree\ (6\degree) in right-ascension (declination). Second row:  component separated and lensing maps of the D56 region from \cite{madhavacheril2019atacama} and \cite{darwish2020atacama}. The cutouts are 5\degree\ on each side and centered at 12.5\degree\ (0\degree) and 17.5\degree\ (-2.5\degree) in right-ascension (declination), respectively. The lensing map has been low-pass filtered with a cutoff at $\ell=200$ to isolate signal-dominated modes (see Figure B2 of \cite{darwish2020atacama}). Bottom left:  the ACT best-fit $TE$ power spectrum (from the ACT likelihood) together with binned data points (C20).  Bottom right: 2D parameter distribution generated from the Markov Chain Monte Carlo (MCMC) parameter chains (A20). }
    \label{fig:data products}
\end{figure}

ACT has operated with multiple generations of cameras, each housed in a cryostat with three separate cryogenic optics tubes holding silicon reimaging lenses.  The earliest data in these releases are from ACT-MBAC, a set of three polarization insensitive detector arrays that were used for the s08--s10 data and observed at 150, 220, and 277\,GHz \citep{swetz/etal:2008}.  In 2013 ACT deployed the polarization sensitive ACTPol receiver, which ultimately included three polarized detector arrays, including a dichroic array, and observed at 98 and 150\,GHz \citep{Thornton2016}.  Most recently, ACT updated to the Advanced ACTPol camera, which was used for the s17--18 data. This camera includes four dichroic detector arrays (three in use at a time), covers close to half the sky, and observes at 30, 40, 98, 150, and 220\,GHz \citep{Henderson2016}, with the 30 and 40\,GHz detectors included from 2020. 

These releases use data from all three epochs, with a primary focus on the s13--s16 data from ACTPol for DR4 and the addition of s17--s18 for DR5. The s13--s16 observations cover seven regions (or ``patches")
of the sky in the 98 and 150\,GHz frequency bands in both temperature and polarization and are discussed in detail in A20 and summarized in \S \ref{sec: splits} in this paper. Data from all three ACT receiver generations, including the additional s17--s18 data, are used in combination with {\sl Planck} to create high-resolution, large-area, and deep ``coadded" maps, presented in N20 and summarized in \S \ref{sec: coadded} and \S\ref{sec: DR5 coadd}. 

\section{Data Products\label{sec: data}}
The data products included in this release are outlined in Table \ref{table:data_products}, and comprise over 5\,TB of data in total (4.5\,TB are simulations). The data set, without the simulations, are publicly available on LAMBDA while simulations are housed on NERSC. Instructions on how to access NERSC are available on the ACT data products page on LAMBDA. Questions about these products can be directed to the ACT data help desk.\footnote{The help desk can be reached at  \href{mailto:act_data_support@googlegroups.com}{act\_data\_support@googlegroups.com} and more information is available on LAMBDA.}

The majority of the maps are released in the Flexible Image Transport System (FITS, see e.g. \cite{fits}) format; the FITS system is widely used in the astronomy community. As described in A20, the $Q$ and $U$ Stokes polarization components of all of the maps are defined using the IAU convention, specified in the FITS file header by the keyword {\texttt{POLCCONV=IAU}}. Earlier pre-DR4 ACT maps were released following the {\texttt{COSMO}} convention \citep{healpix_primer}.

These maps are all released as CAR, or Plate Carrée, maps with rectangular pixelization and x and y axes aligned with right-ascension and declination, respectively.   This is distinct from the HEALPix\footnote{\url{https://healpix.sourceforge.io/} \citep{Gorski_2005}} format commonly used for CMB data.  The CAR projection is a rectangular projection where each pixel represents a constant interval in right-ascension and declination.  This results in pixel areas that depend on latitude, with the largest areas at the equator. Compared to HEALPix, the CAR pixelization has the advantage of simpler and hence faster coordinate calculations; simpler extraction and representation of sub-maps; arbitrary resolution, and a simple and exact pixel window. Because these CAR maps are stored as 2D arrays, they can also be displayed as is, without needing to transform them into another projection, as is the norm with HEALPix. On the other hand, CAR has the disadvantage of increasingly oversampling the maps as one moves away from the equator.\footnote{It is sometimes thought
that CAR maps inherently make assumptions about the sky being flat, and that HEALPix is needed for a curved-sky analysis. This is not the case. The question of flat-sky vs. curved-sky  is not connected to the projection used, and instead depends on how one analyzes the data in the map, e.g. whether one uses Fourier or spherical harmonic transforms. \texttt{pixell} (a software package discussed in \S \ref{sec: nb2}) fully supports curved-sky operations on CAR maps, though it allows one to use the flat-sky approximation should one want it.} The Jupyter IPython notebooks presented in \S \ref{sec: notebooks} provide tutorials with explanations of how to use these maps.

\begin{figure}[!t]
    \centering
    \includegraphics[width=\textwidth]{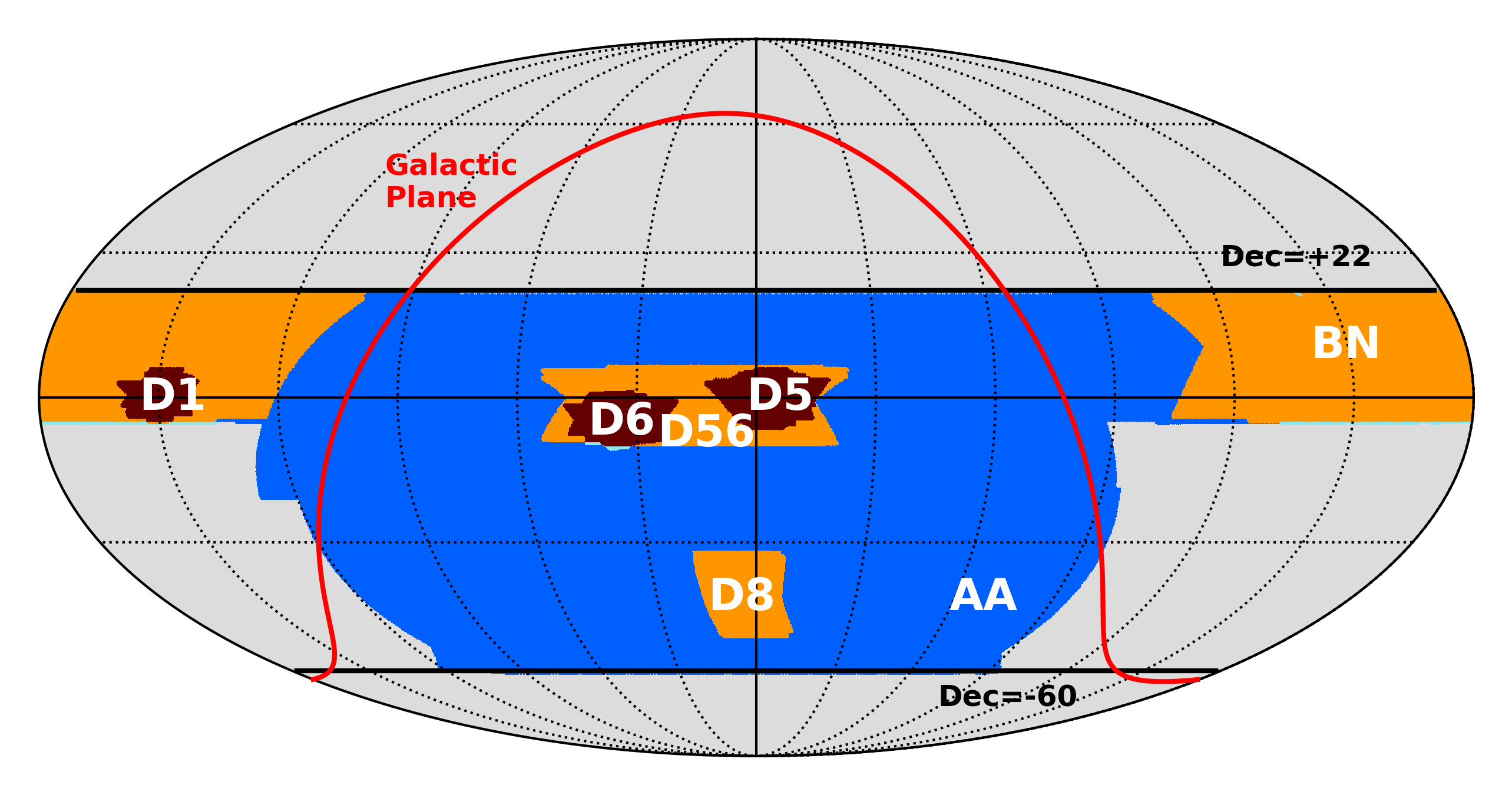}
    \caption{A schematic representation of the observation regions, or patches, in the DR4 frequency maps, as presented in A20. The map is presented in equatorial coordinates, with the D1, D5, and D6 patches in dark red, the D56, D8, and BN patches in orange, and the AA patch in blue. The coadded maps from N20 discussed in \S \ref{sec: coadded} and \S \ref{sec: DR5 coadd} cover the full area between the lines of constant declination, given in degrees. The Galactic plane is overplotted in red for reference. Related figures appear in A20 and C20, in Galactic coordinates.}
    \label{fig: patches}
\end{figure}

\subsection[DR4 Frequency Maps]{DR4 Frequency Maps\footnote{ Maps originally presented in A20.}\label{sec: splits}}
With the exception of the simulations, the largest component of DR4 is $\sim$ 300\,GB worth of CMB maps from the s13--s16 nighttime data. These maps span close to half the sky at 0.5\,arcmin x 0.5\,arcmin resolution and include 94 data sets, each of which contains four FITS files as outlined below. These maps, which are the basis for all of the other ACT DR4 data products, are presented in A20. Section 5.1 of that paper explicitly details the format of the products and files available. 

The maps are grouped according to the observation season, detector array, patch, and frequency channel.  For this release there are seven separate patches which range in size from $\sim$ 100\,deg$^2$ to $\sim$ 17,000\,deg$^2$, shown in Figure \ref{fig: patches}. The release also includes data from the 98 and 150\,GHz channels (depending on detector array) from three detector arrays, referred to as PA1, PA2 and PA3. Precise sky areas for each patch, the different season/detector array/patch/frequency combinations, and the corresponding map-space white noise levels in $\mu \textrm{K}_{\rm CMB}$-arcmin are given in A20 and repeated here in Table \ref{table:maps_stats}.

The map set includes both separately processed subdivisions -- or ``splits" -- of the data as well as coadded maps. For each detector array/frequency/patch/season the release includes four splits, with the exception of the AA patch (named for AdvACT and shown in Figure \ref{fig: patches}), for which there are only two splits. These splits are made by dividing the time-ordered data into non-overlapping, day-long blocks. Each split is evenly assigned these blocks in order to ensure the resulting split maps have similar hit count distributions (A20). Each split map therefore has an independent realization of the noise (see C20 and \S \ref{sec: nb_9}), making them useful for cross correlations. The coadded maps are inverse variance weighted co-additions of the splits, which are not the same as the more complicated multi-array, multi-frequency ``coadds" found in N20.

The splits are discussed in greater detail in A20 and C20 Section 6.1. They are also used for the power spectrum analysis presented in C20 and in \S \ref{sec: spectra}, since the noise in each split is independent. Because these maps are  well-characterized and include separate splits of the data they are optimal for precision cosmology. 

In addition to the CMB maps the release includes a set of other data products that can be used for relevant analyses. In particular, the following data are available:
\begin{itemize}
    \item Source-free maps:  these contain $I$, $Q$, and $U$ Stokes components, using the IAU polarization convention, and are in units of $\mu$K (differential CMB temperature units, hereafter $\mu \textrm{K}_{\rm CMB} $). Some sources are subtracted in the time domain during map-making, while others are subtracted at the map-level (see Section 4.5 of A20 for details on the point source subtraction). 
    \item Source maps:  these contain the point-source signal removed from the full maps in units of $\mu \textrm{K}_{\rm CMB} $. The sum of the source-free and source maps in a given data set gives the 
    total sky map. 
    \item Inverse-variance (``ivar") maps: akin to traditional hit count maps, these maps contain an estimate of the inverse-variance \textit{per pixel} of the frequency maps. They are used (for example) to construct masks for the source-free maps as demonstrated in notebook 8 (\S \ref{sec: nb8}). However, unlike traditional hit count maps, these include the variance contribution from detector noise as is necessary for data operations in map-space. Although in principle the polarization components at each pixel are correlated, DR4 makes available only the ``$II$" element of the inverse-covariance matrix, for two reasons: first, in practice the matrix is diagonal, and second, the ``$QQ$" and ``$UU$" elements are $\sim$ half that of the ``$II$" element (A20). The release provides the ``ivar" maps in units of $(\mu \textrm{K}_{\rm CMB})^{-2}$. 
    \item Cross-linking maps: cross-linking is the degree to which a pixel is uniformly covered in \textit{scan direction}. A pixel is well cross-linked if the telescope scans across it in multiple different directions (e.g. up/down, left/right, diagonally, etc.). Conversely, a pixel is not cross-linked if the telescope always takes the same path across it. The cross linking maps, like the maps themselves, include three components. The first ($T$) component is the inverse variance map; the second ($Q$) component measures the angle between the scanning direction and north; the third ($U$) records whether a given scan was on the rising or setting side of the sky (A20). For example, the equivalent to the polarization fraction of the cross-linking maps gives the cross-linking fraction, where 0\% means uniform coverage in scan direction, and 100\% means only one scan direction covers this pixel (C20).
\end{itemize}

\begin{table}
\movetableright=-.8in
\centering
\begin{tabular}{l|cccc}
\multicolumn{5}{c}{\normalsize{DR4 Frequency Maps: Summary of Patch Sizes and Map White-noise Levels}}\\\addlinespace[7pt]
\toprule
\multirow{3}{*}{\textbf{Map Patch and Area}}& \multicolumn{4}{c}{\textbf{Season, Detector Array, and Frequency}} \\
& s13 & s14 & s15 & s16\\
 & PA1$_{150}$ & PA1$_{150}$/PA2$_{150}$ & PA1$_{150}$/PA2$_{150}$/PA3$_{98/150}$ & PA2$_{150}$/PA3$_{98/150}$ \\
\midrule
D1 (131\,deg$^2$) & 19 & --- & --- & --- \\
& & & &\\
D5 (157\,deg$^2$) & 16 & ---  & --- & --- \\
& & & &\\
D6 (135\,deg$^2$) & 13 & --- & --- & --- \\
& & & &\\
D56 (834\,deg$^2$) & --- & 32/21 & 33/22/18/29 & --- \\
& & & &\\
D8 (248\,deg$^2$) & --- & --- & 42/22/20/29 & --- \\
& & & &\\
BN (3,157\,deg$^2$) & --- & --- & 77/41/34/49 & --- \\
& & & &\\
AA (17,044\,deg$^2$) & --- & --- & --- & 73/79/120 \\
\bottomrule
\end{tabular}
\caption{\label{table:maps_stats} White noise levels in $\mu \textrm{K}_{\rm CMB}$-arcmin for the individual season/detector array/frequency/patch maps. Observation patches correspond to those presented in Figure \ref{fig: patches}. The area on the sky of each patch is also listed. ``PA" refers to ``polarization array," or polarization-sensitive detector array. Subscripts indicate frequency band; PA3 is dichroic, with detectors observing the sky at both 98 and 150\,GHz. This reproduces Table 2 of A20, with white noise data rounded to two significant figures.}
\end{table}

\subsection{DR4 Derived Maps}\label{sec: Derived}
The DR4 frequency maps from \S \ref{sec: splits} can be used to form ``derived maps." These products include coadded maps in temperature and polarization from N20, component separated maps from \cite{madhavacheril2019atacama}, and lensing maps from \cite{darwish2020atacama}.  These products are all outlined in Table \ref{table:data_products}.

\subsubsection[DR4 Coadded Maps]{DR4 Coadded Maps\footnote{ Maps originally presented in N20.}}\label{sec: coadded}
While the individual map splits together span a large sky fraction, they do so in small patches and with the noise characteristics of individual seasons and detector arrays. To complement these smaller, noisier maps DR4 includes full ACT-observed sky coadded maps in the 98 and 150\,GHz frequency channels,  as described in N20. One of these maps is shown in the top panel of Figure \ref{fig:data products}.  These maps combine ACT data from s08--s16 with data from {\sl Planck}, preserving ACT's high-resolution and sensitivity to small scales while adding {\sl Planck}'s sensitivity to large angular scales. The procedure results in individual temperature and polarization maps covering $\sim$ 18,000\,deg$^2$ ($f_{\rm sky} \sim 0.44)$ at a resolution of 0.5\,arcmin x 0.5\,arcmin. Since these maps are deeper, and because quoted sky area is a function of noise level, these coadded maps cover marginally more area than the maps in \S \ref{sec: splits}. The full-coadded depth is $\leq$ 20 $\mu \textrm{K}_{\rm CMB}$-arcmin over $\sim$ 2,600\,deg$^2$ (see Figure 14 of N20). 
See \S \ref{sec: DR5 coadd} for the corresponding improvements in map-depth when s17--s18 data from the AdvACT survey are included in DR5.

DR4 includes coadded maps with and without point sources, as ACT-only and ACT+{\sl Planck}, and in the 98 and 150\,GHz channels. In addition to the maps themselves, DR4 includes associated inverse-variance maps, which can be used to characterize the noise in the maps.

The coadded maps are not recommended for precision cosmology as they lack map splits needed to generate noise simulations or run null tests; the DR4 frequency maps in \S \ref{sec: splits} are more appropriate. \S \ref{sec: DR5 coadd} describes applications that are  well suited for the coadded maps.

\subsubsection[DR4 Component Separated Maps]{DR4 Component Separated Maps\footnote{ Maps presented in \cite{madhavacheril2019atacama}.}\label{sec: compsep maps}}
DR4 includes component separated maps of the CMB+kSZ black-body component, and the thermal Sunyaev-Zeldovich Compton-$y$ parameter, presented in \cite{madhavacheril2019atacama}.  These maps combine multi-frequency data from ACT and the {\sl Planck} satellite to construct arcminute resolution, component separated maps that cover a wide area ($\sim$ 2,100\,deg$^2$). The CMB+kSZ and Compton-$y$ maps are provided for the BN and D56 regions shown in Figure \ref{fig: patches}.  We also show one of these component separated maps in Figure \ref{fig:data products}.

These maps are constructed from an internal linear combination (ILC) of the ACT maps at 98 and 150\,GHz and the {\sl Planck} maps at 30, 44, 70, 100, 143, 217, 353 and 545\,GHz. The algorithm \citep[see e.g.,][]{ilc_2011} uses the frequency dependence of a signal of interest to create a map that preserves that specific signal while minimizing the overall variance.  This minimization is done in the 2D Fourier domain to adapt to the noise properties of a ground-based experiment like ACT.  Maps that explicitly null foreground contamination from specific components with assumed frequency dependence are also provided; in particular (a) Compton-$y$ maps that null the contribution of a dust/CIB-like spectral energy distribution (useful for checks of systematic contamination in cross-correlations) and (b) CMB+kSZ maps with tSZ deprojected for use in lensing estimators, among others. 
See \cite{madhavacheril2019atacama} for more discussion of these maps, the procedure used to generate them, and the relevant caveats.

\subsubsection[DR4 Lensing Maps]{DR4 Lensing Maps\footnote{Maps presented in \cite{darwish2020atacama}.}\label{sec: lens}}
The DR4 s14--s15 data were also used to construct gravitational lensing maps, as described in \cite{darwish2020atacama}. The maps are available for the D56 and BN regions (see Figure \ref{fig: patches}), in versions with and without {\sl Planck} data. One of the lensing maps is shown in Figure \ref{fig:data products}.  
The lensing maps are produced by first coadding the maps at 98 and 150\,GHz and over two seasons (s14--s15) to produce input maps for the lensing estimator. The maps are then convolved to a common beam, inpainted around bright sources in order to account for foreground biases, and cleaned to reduce the effect of ground contamination.  These maps are then passed through the lensing reconstruction pipeline which uses a minimum variance quadratic estimator.  The end result of this process is a set of maps that cover $\sim$ 2,100\,deg$^2$ on the sky and are signal dominated on large scales.  They also overlap with a number of optical surveys which enables them to be used for cross correlations with BOSS CMASS galaxies, for example \citep[see][]{darwish2020atacama}.  DR4 also includes lensing maps with the thermal Sunyaev-Zel'dovich (tSZ) signal removed, following the procedure outlined in \cite{darwish2020atacama} (building on the method from \cite{Madhavacheril_2018}), which has a minimal effect on the lensing signal-to-noise ratio.

\newpage

\subsection{DR4 Ancillary Products}

The ACT map-based and cosmological data products require several secondary inputs -- or ancillary data products -- to support specific analyses, including masks and beams. The masks include point source and extended source spatial masks produced by the ACT map-making pipeline (A20), as well as masks of Galactic foregrounds produced by {\sl Planck} \citep[C20,][]{Planck-overview:2018}. The combined spatial masks serve as inputs to the ACT power spectrum pipeline (C20). In addition to the point source masks, DR4 includes a set of masks based on the multifrequency cross-linking and inverse-variance maps (C20), labeled ``footprint" masks in Table \ref{table:data_products}. It also includes the spatial masks used in \cite{darwish2020atacama} to make lensing maps, and in \cite{madhavacheril2019atacama} to make component separated maps.

For many analyses, correct treatment of the ACT maps rely on an accurate understanding of the beam. A20 provides a detailed description of the beam model as well as the beam products included in DR4. There are four main differences between the various beam products:
\begin{itemize}
    \item Real-space or harmonic space: we include both the real-space azimuthally-averaged beam profiles and the harmonic-space beam transfer functions.
    \item ``Instantaneous" vs ``jitter": the jitter beams include effects of pointing-error variance while the instantaneous beams do not. The jitter beams are recommended for map-domain analyses.
    \item Corrections: each beam is also provided with and without a Rayleigh-Jeans-to-CMB spectral correction for the ACT bandpasses.
    \item Season/detector array/frequency(/patch): the release includes beams for each of the maps and the file names indicate which map corresponds to which beam. Note that only the jitter beams contain patch flags, and so these are not needed for the instantaneous beams.
\end{itemize}

The maps have not been corrected for the smoothing introduced by the beam. Therefore, users must properly account for the beam in their analysis. For map-based analyses, such as comparisons between data sets, users will need to convolve the secondary data set with the ACT beam. Similarly, users can correct for beam effects in harmonic space by dividing out the beam transfer function \citep{bond_1987}. In addition to the beams themselves, the beam leakage between temperature and polarization is also characterized. DR4 includes the $T$-$E$ and $T$-$B$ harmonic-space beam leakages for both the beam core and the polarized sidelobes, as discussed in A20. The beams for the component separated maps are also provided; the component separated maps themselves (\S \ref{sec: compsep maps}) are already convolved with these beams. Finally, DR4 makes available the harmonic-space mapping transfer function associated with the ground-pickup filter used in C20.

\subsection{DR4 Power Spectra}\label{sec: spectra}
DR4 includes the $TT$, $TE$ and $EE$ power spectra described in C20. The multi-frequency spectra at 98 and 150\,GHz are derived from the DR4 frequency maps. We show one of these spectra in Figure \ref{fig:data products}. These spectra are combined with intensity spectra from s08--s10 to produce CMB-only $TT$, $TE$ and $EE$ spectra. More information relevant to these products is found in A20 and C20.

In addition to this full suite of DR4 spectra, the release includes the birefringence spectra from \cite{Namikawa2020}, based on s14 and s15 of the ACT polarization data. Lensed and delensed spectra from the D56 region are presented in \cite{hanetal2020} and \S \ref{sec: delense}.

\subsection{DR4 Likelihoods and Cosmological Parameters}
The maps discussed in the previous section are used to produce a number of products for cosmological analyses including two likelihoods. These consist of a multi-frequency likelihood, and a CMB-only likelihood. The CMB-only likelihood, which does not require sampling over foreground parameters and is simplest for use in cosmological parameter estimation, has versions both in Fortran 90 and in Python.\footnote{\url{https://github.com/ACTCollaboration/pyactlike}} As discussed in A20 and C20, the likelihoods are used to generate Markov Chain Monte Carlo (MCMC) chains to derive constraints on cosmological parameters.  A set of these chains has also been included in the release. A 2D parameter distribution plot generated from one of these chains is shown in the bottom row of Figure \ref{fig:data products}. 

\subsection{DR4 Simulations\label{sec: sim}}
Simulations are key tools in the validation of the ACT data products. 
They are used throughout the ACT analyses in order to calculate transfer functions and covariance matrices, as well as to perform consistency tests. Due to the quantity of data associated with these simulations (90\% of the DR4 disk space), they are not directly available on the public LAMBDA server; instead, they are accessible through NERSC. However, the instructions for access to the simulations, as well as the location at NERSC, are posted on the same LAMBDA webpage as the other data products. The DR4 simulation products (see Table \ref{table:data_products}) are either maps or the power spectra of various components that comprise those maps, and are shared amongst the DR4 product pipelines. In addition, \cite{madhavacheril2019atacama} and \cite{darwish2020atacama} manipulate simulated maps as inputs to, or tests of, other analyses. 

The ``base level" simulation product used throughout DR4 are observed-sky simulations. Their goal is to efficiently replicate the raw ``telescope-observed" sky, including lensing, foregrounds, and noise, analogous to the DR4 frequency maps in \ref{sec: splits}. They have several layers: 

\begin{itemize}
    \item Lensed, Gaussian realizations of the primary CMB with cosmological parameters set equal to the {\sl Planck} 2018 results \citep{Planck-cosmology:2018}. The release provides the \texttt{lensedcmb\_alms}, which are the spherical harmonics of the lensed CMB simulations.. Since these are given in $a_{\ell m}$ format, they must be projected into map space. The release also provides the \texttt{lensing\_phi\_alms} that generate the lensing potential present in the \texttt{lensedcmb\_alms}.
    \item Gaussian realizations of extragalactic foregrounds such as tSZ and radio galaxies.\footnote{Though not formally part of DR4, the foreground power spectrum is available from the \texttt{actsims} repository as \texttt{fg.dat}; see \url{https://github.com/ACTCollaboration/actsims}.}
    \item Gaussian realizations of the map-based noise in each patch, according to the prescription outlined in C20 and discussed in depth in \ref{sec: nb_9}. The release provides ``square-root covariance" matrices that facilitate fast sampling of the Gaussian noise field. These matrices are represented in the same basis as the noise covariance, with eigenvalues equal to the square-root of the noise covariance eigenvalues. See e.g. Appendix B of C20 for an explanation.
\end{itemize}

As the noise is the most technical of these three layers, we provide an example of a ``square-root covariance" matrix and associated usage code in one of our notebooks (\S \ref{sec: nb_9}).\footnote{The DR4 matrices match the ones used in C20. However, due to a version update in September 2019, users will find percent level differences when using the most recent version of \texttt{pixell}.  If users would like to exactly recreate the noise simulations as utilized in C20, they should install a version of \texttt{pixell} released prior to September 17th, 2019. This version is available at \url{https://github.com/simonsobs/pixell/pull/77}}  This base set of ``telescope-observed sky" simulations are direct inputs to C20, \cite{madhavacheril2019atacama}, and \cite{darwish2020atacama}. 

The base simulations are also accompanied by 560 simulations of the component separated maps discussed in \S \ref{sec: compsep maps}.  These simulations mirror the format of the component separated maps and can be used to generate covariance matrices and characterize the data presented in \cite{madhavacheril2019atacama}. The foreground-cleaned observed-sky simulations are passed as inputs to 511 simulations of the lensing-$\kappa$ maps, as discussed in \S \ref{sec: lens}. \cite{darwish2020atacama} then validates the pipeline by cross-correlating the reconstruction with the input lensing fields and finds good agreement.

\subsection[DR4 Delensing Products]{DR4 Delensing Products\footnote{Full data products presented in \cite{hanetal2020}.}\label{sec: delense}}
\cite{hanetal2020} use internally reconstructed lensing maps of the D56 region to produce delensed power spectra at 98 and 150\,GHz. From this, they obtain constraints on $\Lambda$CDM cosmological parameters using the power spectrum of the delensed CMB. 

This release includes the lensed and delensed spectra from D56, the delensing likelihood, and the products derived from the likelihood (the theory curves, the best-fit parameters, the MCMC chains, and the parameter-shift covariance matrix). In order to facilitate direct comparison between the lensed and delensed results the release also provides the lensed spectra of just the D56 area computed with this pipeline.

\subsection[DR5 Coadded maps]{DR5 Coadded maps\footnote{ Maps originally presented in N20.}}\label{sec: DR5 coadd}
The DR5 maps follow the same format as those in Sec \ref{sec: Derived} for the DR4 coadded maps, but they additionally include preliminary maps from s17 and s18 data, daytime data, and data from the 220\,GHz channel, as described in N20. Optimally combining data from the ACT-MBAC, ACTPol, and AdvACT surveys as well as {\sl Planck}, the DR5 coadded maps cover the same total $\sim$ 18,000\,deg$^2$ sky footprint as DR4 but with a full-coadded depth of $\leq$ 10 $\mu \textrm{K}_{\rm CMB}$-arcmin over $\sim$ 2,500\,deg$^2$.

These maps are ideal for producing cluster catalogs, such as in \S \ref{sec: catalog}, and are well suited for stacking analyses such as measurements of the tSZ and kSZ effects \citep{schaan2020atacama,amodeo2020atacama,calafut/etal:2021,vavagiakis/etal:2021}, cluster lensing \citep{Madhavacheril_cluster}, point source and cluster studies, and Galactic science. Due to the preliminary nature of the s17--s18 data, the DR5 coadded maps are not recommended for precision cosmology. This includes $TT$/$TE$/$EE$ power spectrum estimates, reconstructed lensing power spectra and cross-correlations, birefringence, primordial bispectra, and component-separated CMB and Compton y-maps. For a discussion of the technical reasons why the s17--s18 data are preliminary, as well as these caveats, see N20 Section 2.2.3 and the publicly-available explanatory supplement.\footnote{\url{https://phy-act1.princeton.edu/public/snaess/actpol/dr5/lambda/readme.html}} Maps with ACT data observed from s17 onwards, that are appropriate for precision cosmology, are expected in the next ACT data release, DR6.

The DR5 maps are presented with and without point sources, as ACT only and ACT+{\sl Planck}, with the ACT nighttime only or night + day data, and in the 98, 150 and 220\,GHz channels. With the exception of the detailed noise maps below, all maps are produced at 0.5\,arcmin x 0.5\,arcmin resolution. The s08--s18 coadded DR5 maps and the s08--16 DR4 maps are interchangeable in analysis pipelines, including the Jupyter IPython notebook in \S \ref{sec: nb2}.

In addition to the maps, the data release includes a range of ancillary data products:
\begin{itemize}
    \item Inverse-variance maps: similar to those presented in \S \ref{sec: splits}, that provide estimates for the non-atmospheric inverse-variance per pixel in the maps.
    \item Detailed noise maps: in addition to per pixel inverse-variance, these maps also present the inverse-variance across 50 multipole bins. Their pixel size is larger than the simple inverse variance maps: 0.5\,deg x 0.5\,deg vs. 0.5\,arcmin x 0.5\,arcmin. They are also presented on a per-array basis. 
    \item Bandpasses: provided for each of the 15 detector arrays, these bandpasses are normalized for comparison. The normalization is detailed in Equation 6.1 of N20 and results in bandpasses with units of $\mu \textrm{K}_{\rm CMB}/(\rm{MJy}/\rm{sr})/$GHz.
    \item Responses: the map response to individual components such as CMB, tSZ, dust and synchrotron. 
    \item Beams: the harmonic-space beam transfer functions for the maps.
\end{itemize}

\subsection[DR5 Cluster Catalog]{DR5 Cluster Catalog\footnote{Catalog presented in \cite{hilton_atacama_2020}.} \label{sec: catalog}}
\cite{hilton_atacama_2020} present the DR5 cluster catalog, produced using the DR5 
coadded maps. The catalog contains more than 4,000 optically confirmed Sunyaev-Zel'dovich clusters making it significantly larger than ACT's previous catalog. The catalog covers a sky area of 13,211\,deg$^2$ and overlaps with a number of deep optical surveys including the Dark Energy Survey and the Sloan Digital Sky Survey. This overlap with optical surveys makes it possible to obtain redshifts for the clusters by cross matching the catalogs and estimating redshift values.  This process results in mass estimates of clusters that span a redshift range of 0.04 to 1.91 with a median redshift of 0.52, and 221 objects at redshift  $z > 1$. 

The catalog includes right-ascension and declination coordinates, mass, redshift, signal-to-noise ratios, and optical richness measurements \citep[see][for a complete list]{hilton_atacama_2020}. Example analyses using the cluster catalog are presented in notebook 4 and \S \ref{sec: nb4}. The GitHub repository associated with the catalog and the related documentation are both publicly available.\footnote{Repository: \url{https://github.com/simonsobs/nemo/}; and documentation: \url{https://nemo-sz.readthedocs.io/en/latest/}}

\section{Jupyter IPython Notebooks\label{sec:notebooks}} We provide a set of Jupyter IPython notebooks to facilitate use of the DR4 and DR5 data products. The notebooks focus on presenting contextual information about the maps, and serve as an introduction to the tools and techniques needed to manipulate them. We give examples of analyses that can be performed and recreate several key plots from the original DR4 and DR5 papers. These notebooks are new products that were developed alongside the data releases in order to make the ACT data accessible and easy to use.

\begin{figure}
    \centering
    \includegraphics[trim=0cm 0cm 0cm 1cm, clip=true, width=.95\linewidth]{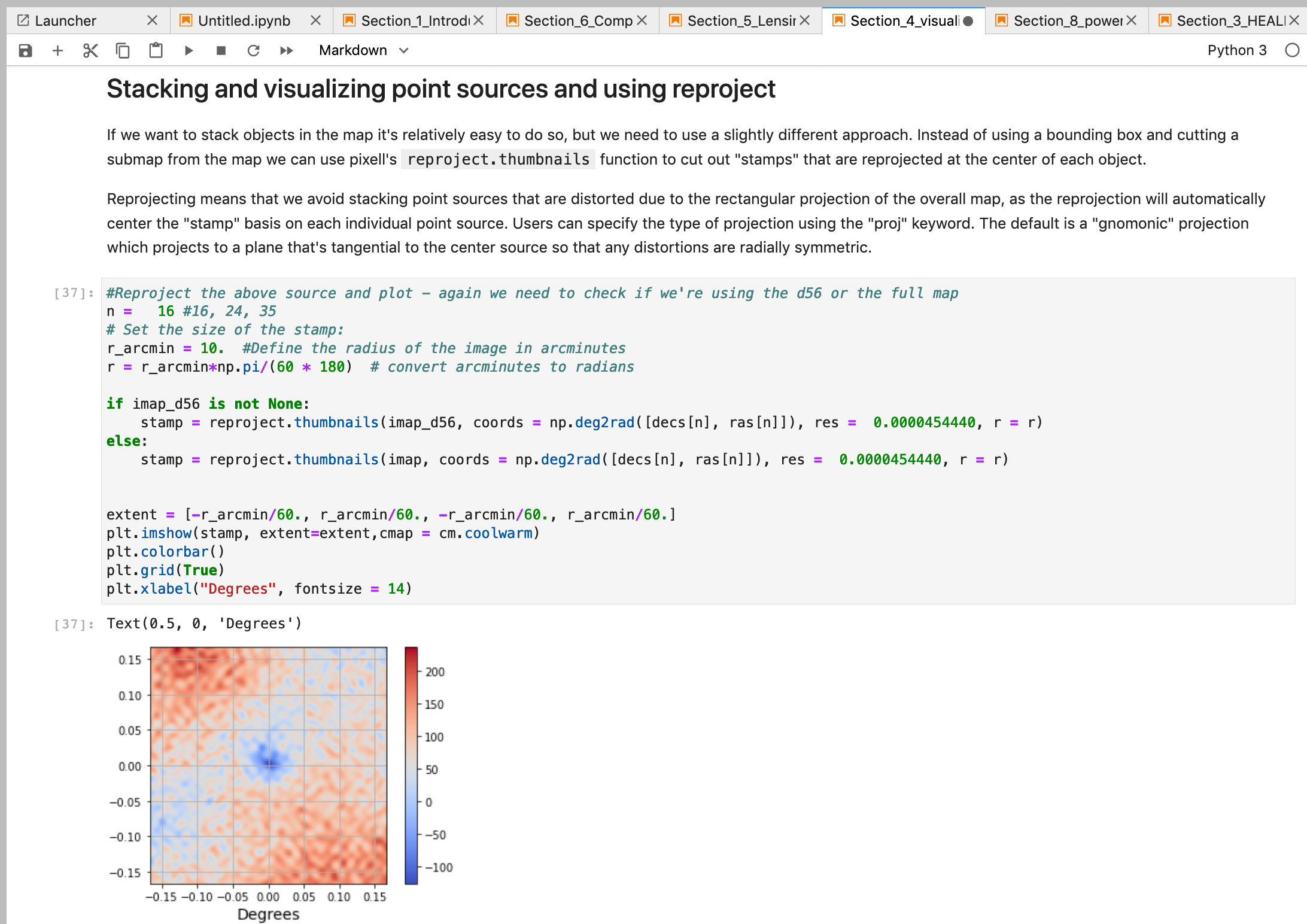}  
    \caption{An example from Notebook 4 that demonstrates how to use a cluster catalog to look at compact sources in the maps.}
    \label{fig:NB4}
\end{figure}

\subsection{Notebook Access}

The set of notebooks discussed in \S \ref{sec: notebooks} are all available via the ACT GitHub repository,\footnote{\url{https://github.com/ACTCollaboration/DR4_DR5_Notebooks}} which includes both the notebooks and instructions for set up and use.  There are two options for the set up; a completely local installation, and installation using a Docker container.\footnote{ \url{https://hub.docker.com/r/actcollaboration/dr4_tutorials}} The local installation involves installing the relevant packages, cloning the GitHub repository and then running scripts that download the necessary data products.  This includes installing a number of less-common packages. The Docker container installation is more straightforward, and comes with all the necessary software pre-installed. 
Users download the container and run the included scripts, which then download the necessary data products. Detailed instructions for set up are included with the notebooks on GitHub. In both cases the time-limiting step is downloading the data, so we provide command-line scripts that pull the data for users and can run in the background on their computer.  Finally, we have set up a help-desk that anyone can contact with questions pertaining to these notebooks.\footnote{The help desk can be reached at \href{mailto:act_notebooks@googlegroups.com}{act\_notebooks@googlegroups.com} and more information is available on LAMBDA.}

\subsection{Notebook Contents}\label{sec: notebooks}
To complement the ACT DR4 and DR5 data releases, we have assembled 12 notebooks. They include explanatory text that focuses both on the data products and the tools to analyze them. The explanations and accompanying code were designed to be accessible to scientists in the astrophysics community regardless of their familiarity with CMB data; however, familiarity with Python and Jupyter IPython notebooks is recommended.  We have also aimed to make these notebooks ``laptop friendly" such that any user can run them locally if they desire.  At some points, where the computational power needed to run certain analyses is particularly high, we have tried to offer alternative solutions so that users can still understand the analysis concepts even if they cannot run the full code locally. 
In this section we provide brief introductions to each notebook, but we do not discuss the full extent of their capabilities. We encourage users to explore the notebooks for themselves. 

\subsubsection{Notebook 1: Overview and Introduction}

This notebook gives an introduction to the two data releases, including an explanation of the CAR map format used in DR4 and DR5.  It is important to understand how this format differs from HEALPix when using these maps. While the notebooks do not need to be completed in order, we encourage users to read through this first one in order to get a sense of the scope and use of the notebooks.  

\subsubsection[Notebook 2:  Guide to Loading and Manipulating the DR4 and DR5 Coadded Maps]{Notebook 2:  Guide to Loading and Manipulating the DR4 and DR5 Coadded Maps\footnote{Maps originally presented in N20.}}\label{sec: nb2} 

This notebook introduces both the tools we use to analyze our maps as well the coadded maps presented in N20 and discussed in \S \ref{sec: coadded} and \S \ref{sec: DR5 coadd}. 
Currently, the majority of public CMB data are presented as HEALPix maps, which differ from the format of the ACT maps.  Due to this difference we walk through the basics of how to interact with maps in the CAR format.  We use the Python library \texttt{pixell},\footnote{\url{https://github.com/simonsobs/pixell}} which is designed for use with CAR maps, and we demonstrate how to load and examine the coadded map.  

\begin{figure}
    \centering
    \includegraphics[width=0.8\textwidth]{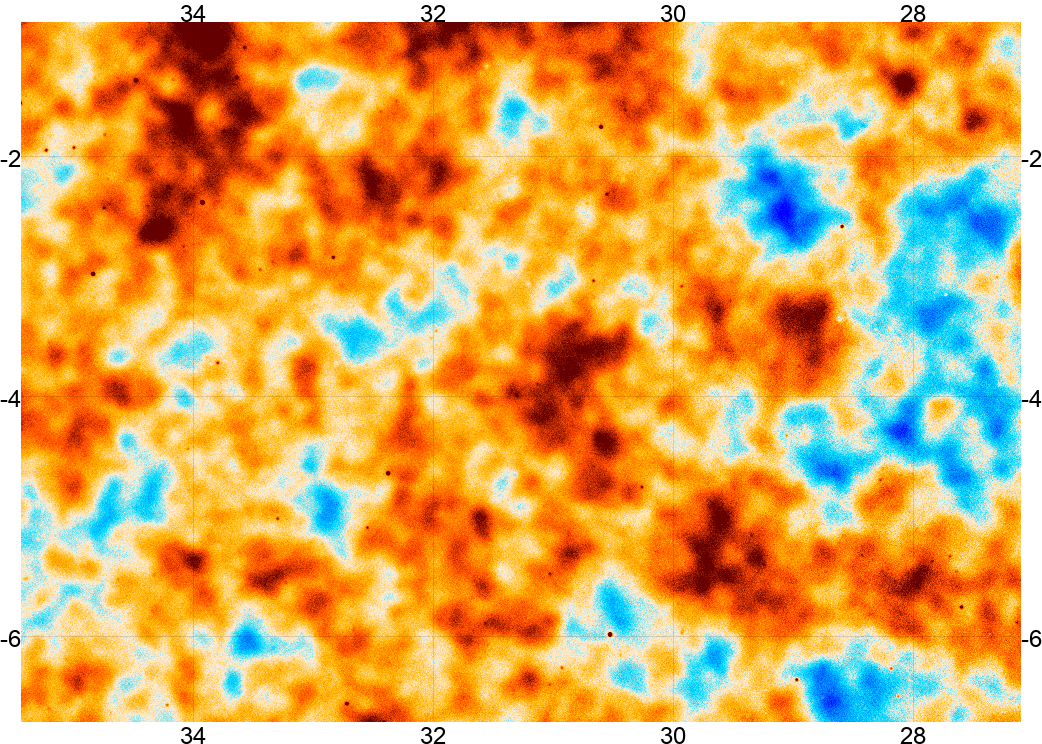}
    \caption{A small section of a full resolution DR5 s08--s18 coadd map (150\,GHz, ACT nighttime + {\sl Planck} data, Stokes $I$). The color scale ranges from -300 $\mu \textrm{\rm K}_{\rm CMB}$ to +300 $\mu \textrm{\rm K}_{\rm CMB}$; the numbers on the x (y) perimeter of the map indicate the right-ascension (declination) in degrees. Individual point sources are visible as small dark red dots at this frequency.}
    \label{fig: coadd_pt_sources}
\end{figure}

The coadded maps are reasonably large data files ($\sim$  5\,GB) and so, while we encourage users to use the full maps if their laptops permit, we also offer a low memory option.  This option involves reading in a low resolution version of the full coadded map as well as a map of a smaller patch of the sky at full resolution.  The low resolution version is created by taking the intensity map from the full coadded map, down-sampling pixels by a factor of four, and then saving the output.  The smaller area map was created with the same input map, but instead of down-sampling, we cut out the D56 region using the D56 footprint included in the DR4 release and then save that as our output.  These two maps combined represent $\sim$ 150 MB of data, compared to the $\sim$ 5 \,GB full resolution data, and so are easier to handle, while still allowing users to appreciate the significant depth and resolution of the ACT coadded maps (see Figure \ref{fig: coadd_pt_sources}).

\subsubsection{Notebook 3: Converting Between CAR and HEALPix Maps}

In many instances it is necessary to convert maps between the CAR and HEALPix format; \texttt{pixell} provides functions that do just that.  We give an example of how \texttt{pixell} can be used to convert a {\sl Planck} map to the same CAR format used for the ACT maps.  Similarly, we offer an example of converting the ACT data products to the HEALPix format if users wish to do so.  

In order to convert between formats, we calculate the spherical harmonics of the given map and then inverse-transform -- or ``reproject" -- those $a_{\ell m}$s onto the new geometry. For that reason, the reprojections are bandwidth limited by the initial harmonic transform.  The reprojection functions can also be used to rotate between different coordinate frames such as from {\sl Planck}'s Galactic frame to ACT's equatorial one.

While these conversions are possible there are certain benefits to using the \texttt{pixell} framework for the ACT maps.  In particular, HEALPix maps do not support analysis using Fourier methods\footnote{Although the Fourier basis is less accurate than the spherical harmonic basis for signals defined on the sphere, for small patches of the sky they are a good and much faster approximation.} and instead must be analysed using spherical harmonics.  Furthermore, \texttt{pixell} simplifies working with smaller and/or arbitrary patches from the full footprint maps. 

\subsubsection[Notebook 4: The DR5 Cluster Catalog and Visualizing Map Objects]{Notebook 4: The DR5 Cluster Catalog and Visualizing Map Objects\footnote{Catalog originally presented in \cite{hilton_atacama_2020}.}}\label{sec: nb4}

One of the advantages of using \texttt{pixell} is that it makes it easier to examine small patches of a map. In this notebook we show how to use the DR5 cluster catalog \citep{hilton_atacama_2020}, together with our maps, to manipulate compact sources and their surrounding regions.  We first examine the cluster catalog and plot the distribution of the objects in 3D space.  We then turn to the maps and extract small patches of the sky around compact sources. We then reproject those cutouts to a plane that is tangential to the sky and centered on the compact source. Switching to this tangential plane, as opposed to the CAR projection used for the full maps, recenters the coordinates such that flat-sky distortions for each cutout are radially-symmetric. This is especially helpful for sources at high declination whose cutouts would otherwise be particularly distorted in the CAR projection. As shown in Figure \ref{fig:NB4}, we extend this analysis by demonstrating how to make stacks of clusters from the reprojected cutouts for use in a variety of analyses.

\begin{figure}
    \centering
    \includegraphics[scale = 0.35]{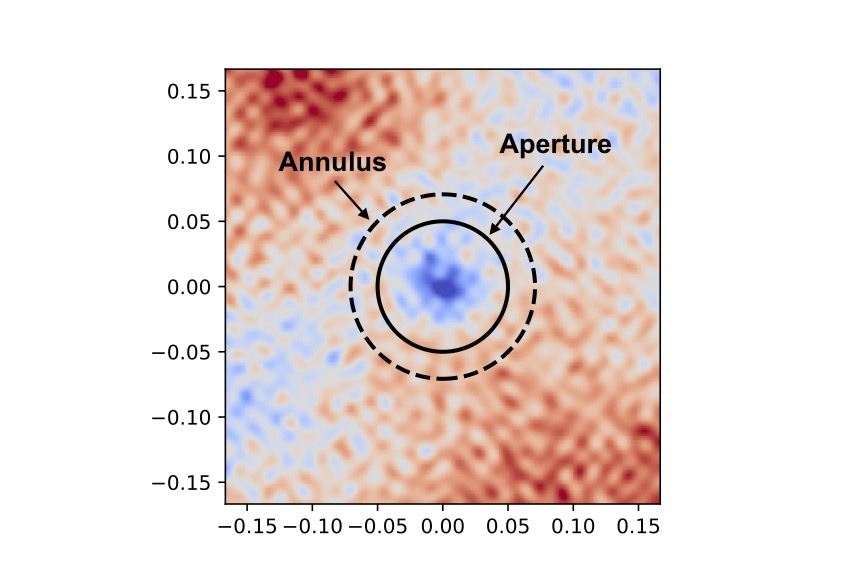}
    \caption{Notebook 4 demonstrates how to apply the filter shown in the image above to stamps of compact sources in the maps. The filtering works by averaging over the disk which is encircled by the solid line, and then subtracting off the average of the equal area annulus, outlined by the dashed line.  The result gives an estimate for the intensity of the compact source, where we have averaged out the background fluctuations. The x and y axes give the scale of the stamp in degrees, with the coordinates centered on the source.}
    \label{fig: AP}
\end{figure}

In order to demonstrate the utility of these cutouts, we show users how to calculate the radially binned profile of the stacked clusters.  Users can then fit profiles to the clusters and use these to study a variety of astronomical objects in our maps.  We also give an example of how users could apply an Aperture Photometry filter, such as the one shown in Figure \ref{fig: AP}, to the stack in order to study the brightness and extent of the clusters that we see. We do this by taking the average of a disk centered on a compact source of radius $r$ and subtracting off the mean of a surrounding annulus of radius $\sqrt{2}r$.  We repeat this for a range of radii and then plot the result against the radius of the inner circle. 

\subsubsection[Notebook 5: Introduction to the Lensing Maps]{Notebook 5: Introduction to the Lensing Maps\label{sec: nb5}\footnote{Lensing maps originally presented in \cite{darwish2020atacama}.}}

We read in a single lensing map \citep{darwish2020atacama} and give an example of how to cross correlate the lensing-$\kappa$ signal with a galaxy density map. This example uses CMASS galaxies\footnote{\url{https://data.sdss.org/sas/dr12/boss/lss/galaxy_DR12v5_CMASSLOWZTOT_South.fits.gz}} from the Sloan Digital Sky Survey (SDSS) data release 12 \citep{reid_at_al_2015}. To create a galaxy density map, $g(\hat{n})$, from this catalog we use 
\begin{equation}\label{density}
g(\hat{n}) = \frac{n(\hat{n}) - \tilde n}{\tilde n}, 
\end{equation}
where $n(\hat{n})$ represents the number of galaxies in a specific pixel and $\tilde n$ gives the average number of galaxies per pixel.  

The notebook also explains how to account for masking effects when using the lensing maps.  In order to create the lensing maps, a mask, known as the CMB window ($W_c$), was first applied to the input CMB maps before they were passed into the lensing reconstruction pipeline \citep{darwish2020atacama}. The released DR4 lensing maps are divided by a factor of $W_c^2$, to correct for the loss of power resulting from the original masking. But since we are introducing a new mask applied to the galaxy map (in addition to the original lensing mask) we need to adjust this factor. This means that we must first remove the $1/W_c^2$ factor in the released map by multiplying by $\langle W_c^2\rangle$. From there we correct for any additional masking that is applied to the maps during the cross correlation as follows:

Option 1: If the user is only applying a window to the galaxy map, as is done in this notebook, they would use
\begin{equation}\label{eq: lensing_mask_1}
    C_L \xrightarrow[]{} \frac{1}{\langle W_c^2 W_L\rangle} C_L
\end{equation}

Option 2: If the user would like to apply a mask to both maps, they would instead use 
\begin{equation}\label{eq: lensing_mask_2}
    C_L \xrightarrow[]{} \frac{1}{\langle W_c^2 W_L^2\rangle} C_L
\end{equation}

\begin{figure}
    \centering
    \includegraphics[width = 0.95\textwidth]{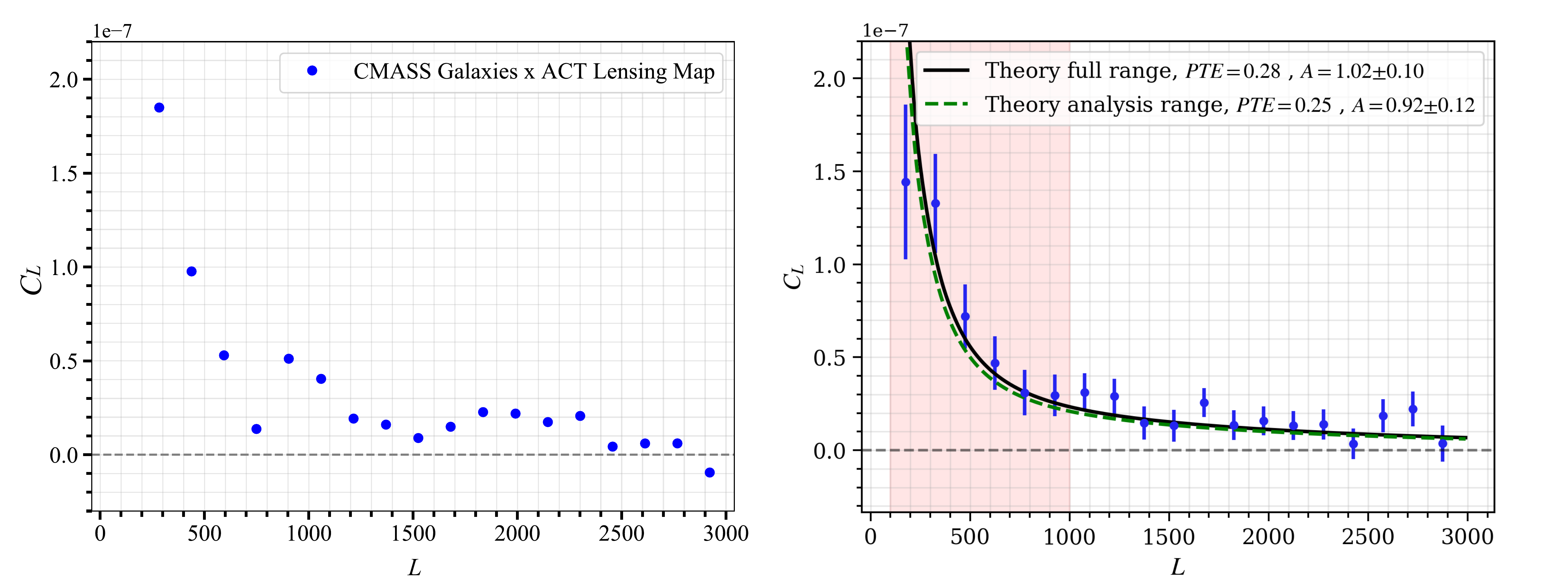}
    \begin{tabular}{c @{\extracolsep{0.03\textwidth}} c @{\extracolsep{0.19\textwidth}} c}
    & Cross correlation using notebook 5 & Figure 5 from \cite{darwish2020atacama}
    \end{tabular}
    \caption{In notebook 5, we introduce users to the lensing maps and demonstrate how to perform a simple cross correlation between these maps and a galaxy overdensity catalog from CMASS, shown on the left. The right plot is from \cite{darwish2020atacama}, which presents the full cross correlation between CMASS galaxies and the lensing map. 
    The differences are due to some simplifying approximations made in the notebook, as noted in the text.
    \label{fig: lensingcross}}
\end{figure}

The resulting cross-spectrum is shown in Figure \ref{fig: lensingcross} where we compare it to the full cross-spectrum from \cite{darwish2020atacama}. The notebook's quick approach differs from the full result, in part, because we do not apply any cuts to the galaxy catalog nor do we apply any weights to individual galaxies when building the galaxy density map. In addition, we use a much smaller map area than used in \cite{darwish2020atacama}. Despite those limitations the results follow a similar overall form and agree in their orders of magnitude. 

\subsubsection[Notebook 6: Guide to the Use of the Component Separated Maps]{Notebook 6: Guide to the Use of the Component Separated Maps\footnote{Maps originally presented in \cite{madhavacheril2019atacama}.}}

This notebook demonstrates how to examine the Compton-$y$ and CMB+kSZ maps included with the component separated maps \citep{madhavacheril2019atacama}.  We also show that it is possible to see galaxy clusters in the Compton-$y$ map by eye and demonstrate how to stack on the locations of these clusters.  This stacking is done using the code we first introduced in notebook 4, 
and the Deep 56 galaxy cluster catalog from ACT DR3 \citep{Hilton_2018}. The same method can be applied to the DR5 catalog.

\subsubsection{Notebook 7:  Harmonic Analysis with CAR maps}

Notebook 7 focuses on how to perform fast harmonic analyses using the CAR maps, walking through both Fourier transforms and spherical harmonic transforms using \texttt{pixell}. 

\begin{figure}[t]
    \centering
    \includegraphics[scale = 0.5]{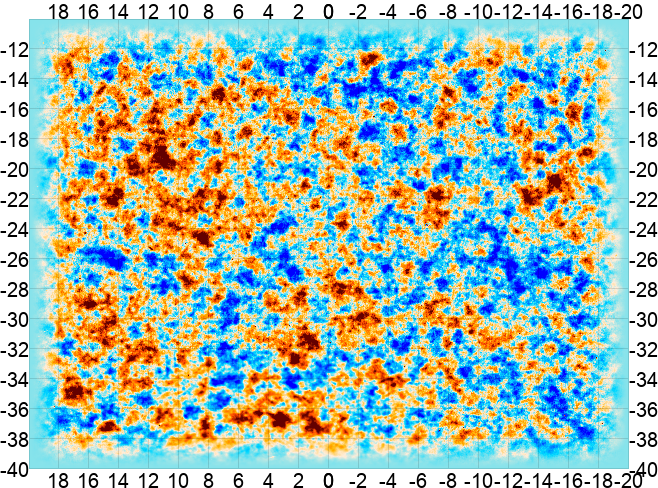}
    \caption{Example of the apodized ACT map from notebook 7 with the x (y) perimeter ticks indicating the right-ascension (declination) in degrees.  This example uses a reduced-resolution version of the s08--s16 DR4 coadd map, which we downgrade in order to make the map easier to handle on a laptop.  We also offer the full map and the routines needed to do the analysis with the full map for users with access to more computational power.}
    \label{fig:act map}
\end{figure}

In order to generate a simple Fourier-based power spectrum we make use of the {\sl Planck} map, projected to the CAR format from notebook 3, and an ACT map from notebook 2.   We apodize the maps to prevent ringing that would occur during the Fourier analysis of the maps due to sharp edges. This process involves smoothly tapering the edges of the maps such that the borders go to zero (an example of an apodized ACT map is shown in Figure \ref{fig:act map}).  Once this apodization has been applied we can Fourier transform the ACT and {\sl Planck} maps, and then cross correlate them. Finally, we compare that cross correlation to the ACT autospectrum.  This approach is fast and makes it easy to swap in other maps for users who are interested in analyzing their own data products.

While the Fourier transform approach works well with smaller maps (where a flat-sky approximation is appropriate), we also provide an example using spherical harmonics.  This approach uses \texttt{pixell} functions to generate $a_{\ell m}$s and \texttt{healpy} functions to produce a power spectrum. 

Both of these approaches are reasonably straightforward to implement and also run quickly on the average laptop.  However, they have a few drawbacks, most noticeably that we fail to properly handle mode-coupling effects from the map mask, as is required for cosmological analyses \citep{Hivon_2002}. This method is also improved upon in the following notebook.

\subsubsection[Notebook 8: Power Spectrum Example using NaWrapper]{Notebook 8: Power Spectrum Example using NaWrapper\label{sec: nb8}}
In this notebook we continue our study of power spectra, but we pivot to using both the individual ACT map splits from A20 and \S \ref{sec: splits}, as well as the \texttt{NaWrapper}\footnote{\url{https://github.com/xzackli/nawrapper}} code base.  This approach offers a few major improvements and much more closely resembles the power spectra analysis that is performed in C20.  However, the improvements also make this notebook one of the most computationally intensive. Ordinarily, we would rarely use code like this in a notebook on a laptop; instead we would submit scripts to a computing cluster to make the process significantly more tractable.  

With that in mind, the code still offers useful insight into the power spectrum analysis process. We begin by generating the correct mask for the region we are considering, in this case the D56 patch. The total mask consists of an apodized footprint combined with the point source mask. The inverse-variance map is then used to correctly weight the pixels. This mask mixes harmonic modes in the naive power spectrum of the map relative to the true power spectrum of the unmasked full sky. This effect is fully captured in a mode-coupling matrix, which links the modes of the naive power spectrum to the true power spectrum \citep{Hivon_2002}. \texttt{NaWrapper}, using the \texttt{NaMaster} code from \cite{Alonso_2019}, is designed to efficiently account for this mode-coupling matrix and recover an estimate for the true power spectrum of an unmasked map. Since this step can be relatively time consuming, we provide the matrix on disk so that users can avoid having to run the computation. Using \texttt{NaWrapper} allows us to improve our estimate of the cosmological power spectrum relative to the previous notebooks (which used \texttt{healpy} and \texttt{pixell}), which were unable to properly decouple mixed harmonic modes induced by the mask. Note that we utilize a slightly different code in the C20 pipeline, but the two codes have been shown to produce the same results (see C20).  

Another difference between this notebook and notebook 7 is that we use four splits of the data (here D56). This allows us to generate six distinct cross spectra, which we average to estimate the power spectrum.  We show two approaches that can be used to generate errors for the power spectra, again providing some of the data products on-disk since the calculations can be difficult to run on the average laptop.

\subsubsection{Notebook 9: Generating Simulations\label{sec: nb_9}}

Simulations play a central role in understanding the ACT DR4 and DR5 products, from the power spectra to the component separation pipeline. In particular, feeding large numbers of map-based simulations through those pipelines enables consistency- and null-tests to be run on spectra, the generation of bin-bin covariance matrices, and assessment of biases in the Compton-$y$ and lensing reconstructions  \citep[C20,][]{madhavacheril2019atacama,darwish2020atacama}. The DR4 release includes a suite of simulation products to complement the data, and in this notebook we walk through an example of how to create simulations of the ACT noise in map space.

The notebook follows Appendix B of C20 to recreate a ``square-root covariance matrix," the on-disk product released for the DR4 noise simulations. As described in C20, the aim is to construct an unbiased noise model of the ACT DR4 maps on a per season, patch, detector array, and frequency basis. Since, in general, the map-domain noise is anisotropic in a given patch (as set by the ACT scan strategy), the 2D Fourier-space power spectrum may contain azimuthally-varying structure.  Starting from the ACT maps, we follow C20 to make estimates of the 2D power spectra (including polarization auto- and cross-spectra), which are assumed to capture all off-diagonal elements of the noise covariance matrix.  In the notebook we apply post-processing steps to remove bias from these estimates, and emerge with the square-root of the noise covariance matrix. Therefore, when written to disk, the ``square-root covariance matrix" enables users to draw fast Gaussian realizations of the underlying anisotropic noise power spectrum. 

\begin{figure}
    \centering
    \includegraphics[width = 0.65\textwidth]{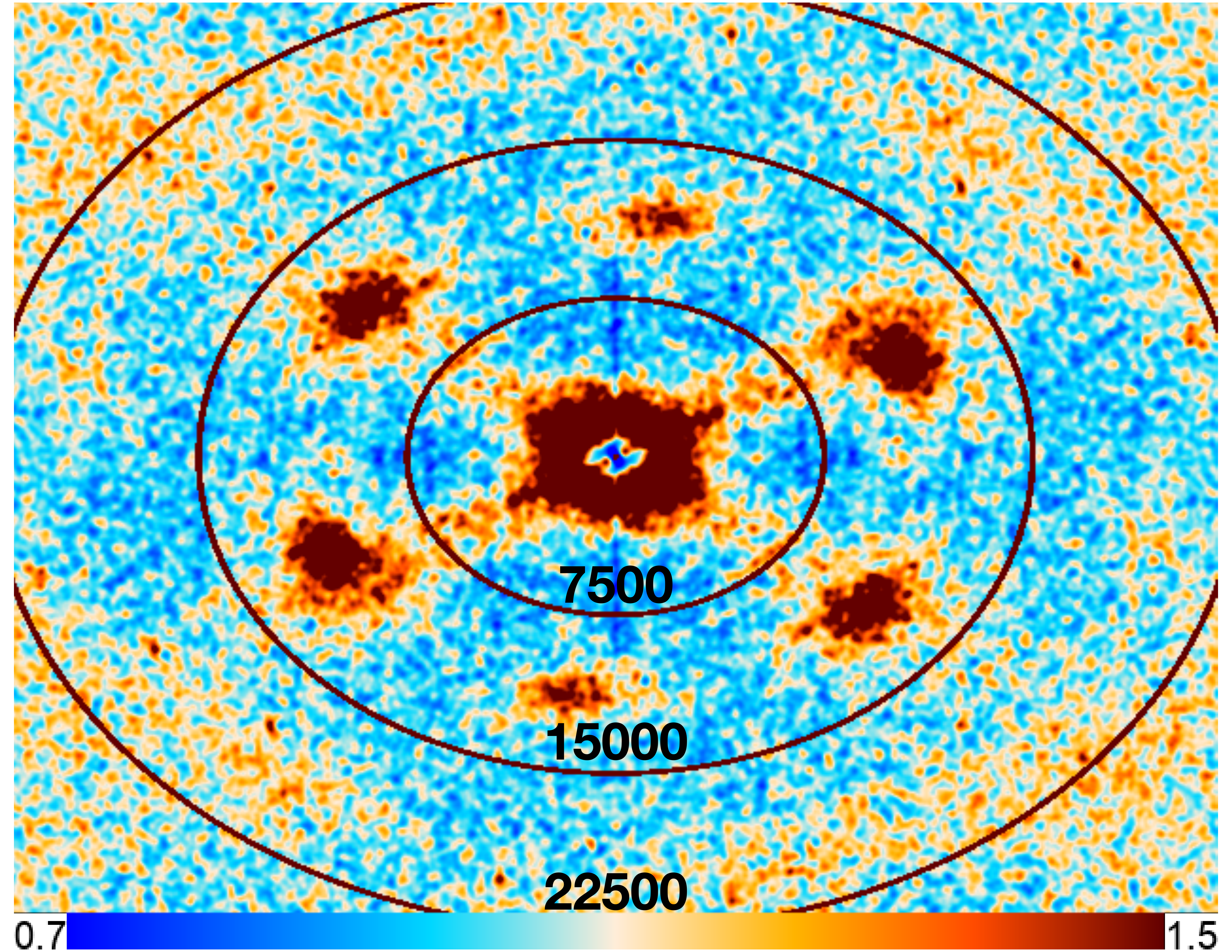}
    \caption{The 2D Fourier-space $I$-$I$ power spectrum for the s13, D6 patch observed by PA1 at 150\,GHz. The concentric circles represent lines of constant $\ell$ given by the labels. The colorbar is unitless: following the noise prescription in Appendix B of C20, this plot represents the power spectrum after it has been radially flattened and subsequently smoothed by a low-pass filter. Therefore this figure shows the relative anisotropy of the 2D noise power, normalized to the average radial profile. The specific anisotropy pattern is a function of atmospheric noise, the detector array geometry, and the ACT scanning strategy.}
    \label{fig: noise_sim}
\end{figure}

In Figure \ref{fig: noise_sim} we show the Stokes $I$-$I$ 2D noise power spectrum for the D6 patch, observed by PA1 in s13, as an intermediate step in the C20 noise simulation prescription. The notebook shows corresponding cross spectra for each of the other polarization combinations. The full ACT map-based simulation pipeline, including signal simulations, is publicly available on GitHub.\footnote{\url{https://github.com/ACTCollaboration/actsims}}

\subsubsection[Notebook 10: Generating E and B maps from I, Q, and U maps]{Notebook 10: Generating $E$ and $B$ maps from $I$, $Q$, and $U$ maps}

In many cases the raw $I$, $Q$, and $U$ maps provided on-disk contain the information needed to perform a given analysis. For example, when studying point sources, Galactic dust, or doing component separation, users may require the Stokes $Q$ and $U$ maps instead of maps of $E$ and $B$ modes. However, many analyses require $E$ and $B$ modes as inputs instead. Because the $E$ and $B$ maps are not included in this release, we provide an example of how users can generate maps of the $E$ and $B$ modes from the Stokes components using \texttt{pixell} functions. The method used is straightforward, but as with many Fourier space operations, decisions such as how users treat bright spots in the maps, or how one apodizes the boundary, can affect the final product. 

\subsubsection{Notebook 11: ACT Likelihood\label{sec: nb11}}

The data release includes the likelihood of the CMB power spectrum measured by ACT from the individual DR4 patch maps discussed in A20 and \S \ref{sec: splits}. The \texttt{pyactlike} package used in this notebook is available on GitHub.\footnote{This code is based on the {\sl WMAP} and ACT Collaboration's likelihood software and is available at \url{https://github.com/ACTCollaboration/pyactlike}} The notebook demonstrates how to plot the best-fitting theoretical power spectra with the $TT$, $TE$ and $EE$ data, compute the likelihood for a given cosmological model, and plot the residuals of the theoretical model compared to the data. We also show how to load one of the MCMC chains generated using the likelihood from A20, and use it to plot 1D and 2D parameter distributions. The chains allow us to visualize the joint distribution of different model parameters; in the notebook we plot the distribution of $H_0$ and $\Omega_m$. Another example of the notebook outputs can be found in the bottom row of Figure \ref{fig:data products}.

\newpage

\subsubsection{Notebook 12: Delensing and Parameter-shift Covariance Matrix}

As discussed in \cite{hanetal2020} and \S \ref{sec: delense}, it is possible to use internally reconstructed lensing maps of the D56 region to undo the effect of CMB lensing in the 98 and 150\,GHz ACT power spectra. This release includes the delensing likelihood for the D56 region, the best-fit parameters obtained from both lensed and delensed spectra, as well as a parameter-shift covariance matrix. The parameter-shift covariance matrix is used to quantify the shift in the best-fit parameters obtained from lensed and delensed spectra. The notebook demonstrates how to read in the various parameters and the covariance matrix, then uses these products to calculate the parameter-shift $\chi^2$ and probability-to-exceed (PTE) values for the given set of lensed and delensed best-fit parameters. 

\section{Conclusion\label{sec: conc}}
The Atacama Cosmology Telescope (ACT) data releases four and five (DR4 and DR5) include maps that span close to half the sky in temperature and polarization at arcminute resolution, along with $\sim$ 2,100\,deg$^2$ of CMB lensing maps, Compton $y$-maps, and more.  DR4 and DR5 also contain data products including beams and masks, signal and noise simulations, power spectra, and likelihoods necessary for a wide range of cosmological analyses. In total, the data releases comprise over 5\,TB of data from multiple generations of ACT cameras, which have been documented in a series of papers.

By including Jupyter IPython notebook tutorials with the data products in these releases, the goal is to make it practical to include ACT data in external analyses. The data products support a multitude of applications, from testing $\Lambda$CDM-cosmology and its extensions, to extragalactic studies such as measurements of the kSZ effect, active galactic nuclei, and dusty star-forming galaxies, to Galactic science. The releases summarized here include data taken through 2018. As the ACT team validate and release data from 2019-2021 as well as additional data from 2017-18, the depth of the temperature and polarization maps will increase. Indeed, as the community prepares for an anticipated wealth of new cosmological data in the next few years (for example, the Vera C. Rubin Observatory \citep{LSST} or the Euclid mission \citep{euclid}), ACT's unique millimeter maps will contribute to the investigation of, among other things, the dark sector, evolution of structure across scales, and other open questions that lie beyond the scope of traditional CMB science.\\

\section*{Acknowledgments}
This  work  was  supported  by  the  U.S.  National  Science Foundation  through  awards  AST-0408698,  AST-0965625,  and  AST-1440226  for  the  ACT  project,  as well as awards PHY-0355328,  PHY-0855887 and PHY-1214379.  Funding was also provided by Princeton University, the  University  of  Pennsylvania, and  a  Canada Foundation for Innovation (CFI) award to UBC. ACT operates in the Parque Astro\'{n}omico Atacama in northern Chile  under  the  auspices  of  the  Comisi on  Nacionalde  Investigac\'{i}on  (CONICYT). The development of multichroic detectors and lenses was supported by NASA grants NNX13AE56G and NNX14AB58G. Detector research at NIST was supported by the NIST Innovations in Measurement Science program. Computations were performed on Cori at NERSC as part of the CMB Community allocation, on the Niagara supercomputer at the SciNet HPC Consortium, and on Feynman  and Tiger at Princeton Research Computing.  SciNet is funded by the CFI under the auspices of Compute Canada, the Government of Ontario, the Ontario Research Fund—Research Excellence, and the University of Toronto. MMK is supported by the NSF Graduate Research Fellowship under Grant No. DGE-1256260. DH, AM, and NS acknowledge support from NSF grant numbers AST-1513618 and AST-1907657. EC acknowledges support from the STFC Ernest Rutherford Fellowship ST/M004856/2 and STFC Consolidated Grant ST/S00033X/1, and from the Horizon 2020 ERC Starting Grant (No 849169). KM acknowledges support from the National Research Foundation of South Africa. ZX acknowledges support from the Gordon and Betty Moore Foundation. ZL, ES and JD are supported through NSF grant AST-1814971.

We gratefully acknowledge the publicly available software packages that were instrumental for this work.  These packages include \texttt{HEALPix} \citep{Gorski_2005} and the python wrapper for \texttt{HEALPix} which is \texttt{healpy} \citep{Healpix1}.  We also made use of \texttt{Astropy},\footnote{\url{http://www.astropy.org}} a community-developed core Python package for Astronomy \citep{astropy:2013, astropy:2018}, \texttt{libsharp} \citep{reinecke/2013} and the \texttt{matplotlib} \citep{Hunter:2007} package.  We used \texttt{pixell}\footnote{\url{https://github.com/simonsobs/pixell}} as well as \texttt{pyactlike}.\footnote{\url{https://github.com/ACTCollaboration/pyactlike}}
We also relied on \texttt{CAMB} \citep{CAMB} as well as \texttt{getdist} \citep{Lewis:2019xzd}. For power spectra examples we used \texttt{NaWrapper}\footnote{\url{https://github.com/xzackli/nawrapper}} which is a wrapper for \texttt{NaMaster} \citep{Alonso_2019}.\footnote{\url{https://github.com/LSSTDESC/NaMaster}} The packages \texttt{numpy} \citep{oliphant2006guide, van2011numpy, harris2020array}, \texttt{pandas} \citep{jeff_reback_2020_3715232, mckinney-proc-scipy-2010} and \texttt{scipy} \citep{2020SciPy-NMeth} were also used throughout this work. We acknowledge the use of the Legacy Archive for Microwave Background Data Analysis (LAMBDA), part of the High Energy Astrophysics Science Archive Center (HEASARC). HEASARC/LAMBDA is a service of the Astrophysics Science Division at the NASA Goddard Space Flight Center. We also 
acknowledge the use of the National Energy Research Scientific Computing Center (NERSC) for hosting the ACT data, and the use of GitHub for hosting our repositories.

\bibliography{main}
\bibliographystyle{aasjournal}

\end{document}